\documentstyle[12pt,epsf]{article}
\def\href#1#2{#2}   
\newif\ifdraft
\draftfalse	  
\reversemarginpar   
\makeatletter
\let\mlabel=\label
\let\adkendequation=\endequation%
\def\endequation{\adkendequation\adklabel\global\@ignoretrue}
\let\adkendeqnarray=\endeqnarray%
\def\endeqnarray{\adkendeqnarray\adklabel\global\@ignoretrue}
\newbox\marglabbox
\def\adklabel{\ifvoid\marglabbox\else\marginpar{\unhbox\marglabbox}\fi}
\def\label#1{\ifdraft\ifmmode%
  \global\setbox\marglabbox=\hbox{\hfill\fbox{\tiny\verb*~#1~}}%
  \else\ifinner\else\marginpar{\hfill\fbox{\tiny\verb*~#1~}}%
  \fi\fi\fi \mlabel{#1}}
\makeatother
\ifdraft%
\fi
\font\twelvebb=msbm12
\font\tenbb=msbm10
\font\sevenbb=msbm7
  \newfam\bbfam
  \def\bb{\fam\bbfam\twelvebb}
  \textfont\bbfam=\twelvebb
  \scriptfont\bbfam=\tenbb
  \scriptscriptfont\bbfam=\sevenbb
\font\twelveeusm=eusm10 scaled 1200
\font\teneusm=eusm10
  \newfam\eusmfam
  
  \textfont\eusmfam=\twelveeusm
  \scriptfont\eusmfam=\teneusm
  \scriptscriptfont\eusmfam=\scriptfont\eusmfam
\font\twelvefrak=eufm10 scaled 1200
\font\tenfrak=eufm10
  \newfam\frakfam
  \def\frak{\fam\frakfam\twelvefrak}
  \textfont\frakfam=\twelvefrak
  \scriptfont\frakfam=\tenfrak
  \scriptscriptfont\frakfam=\scriptfont\frakfam
\def\sqr#1#2{{\vcenter{\hrule height.#2pt
   \hbox{\vrule width.#2pt height#1pt \kern#1pt
      \vrule width.#2pt}
   \hrule height.#2pt}}}

\def\bsqr#1#2{{\vrule width #1pt height#2pt}}
\def\bsquare{{\mathchoice\bsqr66\bsqr66\bsqr33\bsqr33}}
\def\badbreak{\penalty1000}
\newtheorem{theorem}{Theorem}
\newtheorem{lemma}{Lemma}
\newtheorem{definition}{Definition}
\newtheorem{corollary}{Corollary}
\newtheorem{hypothesis}{Hypothesis}
\newenvironment{proof}{{\em Proof.}}{\badbreak$\;\bsquare$\smallskip}
\def\identity{{\bb I}}			    
\def\union{\cup}

\def\rational#1#2{{\mathchoice{\textstyle{#1\over#2}}%
  {\scriptstyle{#1\over#2}}{\scriptscriptstyle{#1\over#2}}{#1/#2}}}
\def\half{\rational12}			    
\def\R{{\bb R}}				    
\def\Z{{\bb Z}}				    
\def\C{{\bb C}}				    
\newcommand{\Dset}{{\frak D}}               
\newcommand{\D}{{\bf D}}                    
\newcommand{\Dr}{{\overline D}}             
\newcommand{\Dc}{{\cal D}}                  
\newcommand{\Dcr}{{\overline \Dc}}          
\newcommand{\Rset}{{\frak R}}               
\newcommand{\Rb}{{\bf R}}                   
\newcommand{\Rbr}{{\overline R}}            
\newcommand{\Tset}{{\frak T}}               
\newcommand{\G}{{\bf G}}                    
\newcommand{\Gr}{{\overline G}}             
\newcommand{\F}{{\bf F}}                    
\newcommand{\Fr}{{\overline F}}             
\newcommand{\Gsl}{{{\frak G}^{sl}}}         
\newcommand{\Hcal}{{\cal H}}                
\newcommand{\psibar}{\overline{\psi}}       
\newcommand{\pbar}{\overline{p}}            
\newcommand{\gmu}{\gamma_{\mu}}             
\newcommand{\gnu}{\gamma_{\nu}}             
\newcommand{\dmunu}{\delta_{\mu,\nu}}       
\newcommand{\gone}{\gamma_{1}}              
\newcommand{\gtwo}{\gamma_{2}}              
\newcommand{\gfive}{\gamma_{5}}             
\newcommand{\cbar}{\overline c}             
\newcommand{\A}{A}                          
\newcommand{\B}{B}                          
\newcommand{\fa}{a}                         
\newcommand{\fb}{b}                         
\newcommand{\fan}{a_n}                      
\newcommand{\fbn}{b_n}                      
\newcommand{\rp}{\rho}                      

\begin{document}

\title{
  Ginsparg-Wilson-L\"{u}scher Symmetry and Ultralocality}

\author{Ivan Horv\'ath\thanks{{\tt ih3p@virginia.edu}} \\[1ex]
  Department of Physics, University of Virginia\\
  Charlottesville, Virginia 22903, U.S.A.}

\date{\today}

\maketitle

\begin{abstract}
  \noindent
  Important recent discoveries suggest that Ginsparg-Wilson-L\"{u}scher
  (GWL) symmetry has analogous dynamical consequences for the theory on 
  the lattice as chiral symmetry does in the continuum. While it is well
  known that inherent property of lattice chiral symmetry is fermion 
  doubling, we show here that inherent property of GWL symmetry 
  is that the infinitesimal symmetry transformation couples fermionic 
  degrees of freedom 
  at arbitrarily large lattice distances (non-ultralocality). 
  The consequences of this result for ultralocality of symmetric actions are 
  discussed.\footnote{This is a generalization and detailed account of 
  Ref.~\cite{Hor98A}, announced as Ref.~[10] in that work. Results 
  discussed here (including proof of Theorem~1) were presented by the author 
  at the workshop VIELAT98, Vienna, Sep 24--26, 1998.} 
\end{abstract}
   
\section{Introduction}

One of the outstanding problems in theoretical particle physics is 
the question of nonperturbative definition of the full Standard Model.
Following the Wilson's work on renormalization group in late sixties 
and early seventies, it became an accepted practice to think of continuum
field theory as a scaling limit of the appropriate model defined on the 
space-time lattice. Quite naturally, then, this approach became a
primary candidate for achieving the goal of defining the theoretical
framework of particle physics nonperturbatively. 

However, lattice field theory became a useful tool in this respect
only to the extent it was able to reflect the important symmetries
encoded in the Standard Model. From the standpoint of principle, 
the only requirement for the lattice--regularized theory is that it 
posesses the critical point with the continuum limit, corresponding to
the target field theory. While the presence of a particular symmetry 
of the target theory at lattice level is not strictly required, 
it is desirable because it makes the lattice theory to resemble 
its target more before the continuum limit is actually taken. 
Thus, the fact that Wilson's formulation of lattice gauge 
theories~\cite{Wil74A} accomodates local gauge invariance exactly, is 
arguably the single most important reason why the lattice approach took
off in the context of high-energy physics.

Including gauge invariance on the lattice marked a nonperturbative 
formulation of QCD with proper gauge dynamics. However, at the same
time, the persistent failure of accomodating chiral symmetries without
fermion doubling kept lattice QCD severly impaired from both theoretical 
and practical point of view, and the lattice definition of the electroweak 
sector was not possible at all. Furthermore, there were serious reasons
to believe that this is actually unavoidable~\cite{Nie81A}.

Sufficiently new ideas with the potential of ending the ``chirally blind'' 
period in lattice field theory only appeared in the early ninetees. 
Starting with the influential paper of Kaplan~\cite{Kap92A}, the subsequent
developments were the variations on the idea that by assigning to every
light degree of freedom additional heavy ones in appropriate manner,
it might be possible to enforce chiral dynamics on the low energy
lattice theory without doubling of fermionic species.
It became soon clear that to achieve strict chirality,
the number of auxiliary degrees of freedom per single light one must be 
infinite. In this respect, {\it domain wall fermions}~\cite{Sha93A}
represent the formulation with finite total number of degrees
of freedom, wherein the violations of chiral symmetry are viewed
as a ``finite volume effect''. The auxiliary degrees of freedom are 
realized by ``extra dimension'' and the chiral limit at fixed number
of light degrees of freedom is achieved as the extension of the extra
dimension becomes large. The domain wall fermion setup is quite natural
for vectorlike theory like lattice QCD, but its use for chiral gauge
theories is not quite clear. Nevertheless, the variation on this 
approach proposed in~\cite{Cre97A} might represent a valid
regularization of the Standard Model. 

The {\it overlap formalism}~\cite{Nar95A} attempts to fully
respect the infinity of additional degrees of freedom. Their effect
is ``sumed up'' into the overlap of ground states of the auxiliary 
finite many body Hamiltonians. This setup is more flexible with
respect to including chiral gauge theories than domain wall fermions
and it may represent a general way of defining these theories 
nonperturbatively. For vectorlike case, Neuberger was able to express
the fermionic partition function given by the overlap formula as
the determinant of the new lattice Dirac operator (Neuberger 
operator)~\cite{Neu98B}. Thus, for vectorlike theory, the overlap 
prescription including auxiliary Hamiltonians can be turned into 
standard fermionic path integral expression with a particular
choice of lattice Dirac kernel. 

Almost in paralell with the above developments, there was a significant
activity on developing further the old idea of perfect action for 
QCD~\cite{Has98C}. Even though defined on the lattice, such action should 
be continuum--like in all dynamical respects, including the dynamical
consequences of chiral symmetry~\cite{Bie96A}. What this formally 
implies for the perfect action is somewhat unclear, but as noted first
by P. Hasenfratz~\cite{Has98C}, for fixed point action (classically 
perfect action) the answer to that question was indirectly given 
long ago by Ginsparg and Wilson~\cite{Gin82A}. In particular, using
renormalization group arguments, Ginsparg and Wilson suggested that 
the correct chiral dynamics can be ensured on the lattice by imposing 
the Ginsparg-Wilson (GW) relation for the lattice Dirac kernel, and 
Hasenfratz has shown that this condition is satisfied by doubler--free
fixed point action.

However, ``perfectness'' is not necessary for GW relation to be satisfied. 
Indeed, in an interesting turn of events, Neuberger has shown that his 
lattice Dirac operator also represents an acceptable solution~\cite{Neu98A}.
It thus turns out that the overlap and domain walls share with fixed point 
action the property of building in the Ginsparg--Wilson lattice chiral 
dynamics. 

L\"uscher put these intriguing developments on more solid formal (and 
also {\ae}sthetic) ground by identifying a symmetry principle behind  
GW relation~\cite{Lus98A}. He proposed a modified chiral transformation 
of lattice fermionic variables, such that invariance with respect to this 
transformation is equivalent to imposing a GW relation. 
This meant that standard field--theoretical language and methods could 
suddenly be used to deal with chirality on the lattice. 
While domain walls and overlap formalism seem rather mysterious
and unnatural to many workers in the field, the new developments
can be sumed up by saying that, instead of standard chiral
symmetry, we need to demand Ginsparg-Wilson-L\"uscher (GWL) symmetry
and to study its field--theoretical consequences. 
The crucial element here is the fact that, while GWL symmetry ensures 
appropriate continuum--like chiral dynamics~\cite{Gin82A,Has98B,Cha98A},
fermion doubling is not a necessity. As expected and hoped for, it now 
appears that (at least U(1)) chiral gauge theories can also be constructed 
based on the fermionic actions with GWL symmetry~\cite{Lus98B}.

The importance of the above formal developments also lies in the fact
that we can now talk in general about the set of actions with 
GWL symmetry (GW actions), 
to study their common properties, to identify additional characteristics 
that could usefully differentiate between them, to identify new explicit
solutions and so on. It is possible that ultimately it will turn out that 
using domain wall fermions, Neuberger operator, or some truncated perfect 
action will be the most practical way to include chiral dynamics in 
lattice QCD. Nevertheless, the field-theoretical language of GWL 
symmetry is very appealing and these are virtually unexplored territories 
with high potential for a surprising result.

In this paper we will study generalized version of original L\"{u}scher 
transformations~\cite{Lus98A} in the context of lattice Dirac kernels
that are local, respect symmetries of the hypercubic lattice, are gauge 
invariant, and posess the correct classical continuum limit. 
Unconventional feature of L\"{u}scher transformations is that their nature 
depends on the dynamics governing the fermionic theory under consideration. 
We show that if the dynamics is invariant, then the infinitesimal
symmetry operation requires rearrangement of infinitely many degrees 
of freedom for every fermionic variable on 
unrestricted lattice. Stated equivalently, the transformation couples 
fermionic variables at arbitrarily large lattice distances 
(non-ultralocality). This means that ensuring GWL symmetry requires a 
delicate collective process involving cooperation of many (perhaps all) 
fermionic degrees of freedom contained in the system. 

Note that this is the same kind of qualitative feature that is present
when we enforce chiral dynamics through domain walls in extra dimension.
When the infinity of additional degrees of freedom that helped to arrange
for chirality are integrated out and Neuberger operator arises, that 
operator (and L\"{u}scher symmetry transformation) couples variables at 
arbitrarily large lattice distances. Our result shows that this is an 
allways--present property of GWL symmetry in the context of acceptable
lattice Dirac operators.

The above conclusion has important implications for GW actions themselves.
In particular, it implies non--ultralocality for the subset of GW operators,
specified in Ref.~\cite{Hor98A} (see also ${\rm footnote}^{\,{\rm 4}}$). 
While this subset is very relevant for practical purposes, the statement is 
most likely true in the most general case as well if one insists explicitly
that the theory be doubler--free. From this point of view, 
we can refer to the theorem on the absence of ultralocal symmetry 
transformations presented here as {\em weak theorem on non-ultralocality}. 
Hypothesis about strict absence of ultralocal doubler--free GW actions 
({\em strong theorem on non-ultralocality}) still awaits its proof. These
issues will be discussed in a separate subsection.

\section{Generalized L\"{u}scher Transformations}

Our main interest in this paper is to study infinitesimal linear 
transformations of the type first proposed by L\"{u}scher \cite{Lus98A}.

\subsection{General Algebraic Structure}

Consider a $d$-dimensional hypercubic lattice (finite or infinite), 
where $d$ is an even integer. Let $\psi,\psibar^T$ are vectors of 
fermionic variables living on lattice sites with the usual 
spin-gauge-flavour structure. Let further $\D,\Rb$ be arbitrary
matrices acting in the corresponding linear space.
To every such pair $(\D,\Rb)$ we assign a one--parameter family of 
infinitesimal transformations
\begin{equation}
   \psi \;\longrightarrow\; \psi + i\theta\gfive (\identity-\Rb\D) \psi
   \qquad\qquad
   \psibar \;\longrightarrow\; \psibar +
        \psibar\, i\theta (\identity-\D\Rb)\gfive\;,
   \label{eq:10}  
\end{equation}
and call them generalized L\"{u}scher transformations. They were
considered for example in Ref.~\cite{Tin98B} for the case when $\Rb$ is
trivial in spinor space and Hermitian. Here we will not make such 
restriction. 

Interesting subset of generalized L\"{u}scher transformations is 
represented by those pairs $(\D,\Rb)$, for which the transformation does 
not change the expression $\psibar\D\psi$ (``fermionic action''). 
The change is given by
\begin{displaymath}
   \delta(\psibar\D\psi)\;=\; \psibar\D\delta\psi + \delta\psibar\D\psi 
            \;=\; i\theta\,\psibar\,\Bigl(\; \{\D,\gfive\} 
                  - \D\{\Rb,\gfive\}\D\;\Bigr)\,\psi\;,
\end{displaymath}
where $\{,\}$ denotes the anticommutator, and vanishes only if 
\begin{equation}
    \{\D,\gfive\} \;=\; \D\{\Rb,\gfive\}\D \qquad\;
    \mbox{or} \qquad\;
    \{ \D^{-1},\gfive \} \;=\; \{ \Rb,\gfive \} \;.
    \label{eq:20}
\end{equation}
We note that the first form of condition (\ref{eq:20}) is fundamental and 
the second one is equivalent to it if the inverse of $\D$ can be meaningfully 
defined. For such $\D$, it can also be written in equivalent explicit form
\begin{equation}
   \Rb \;=\; \D^{-1} + \F\qquad\qquad  \{\F,\gfive\}=0\;,
   \label{eq:30}
\end{equation}
with some arbitrary chirally symmetric $\F$.

\subsection{Physically Relevant Restriction}

We now specify three restrictions that will be used to define the subset 
of generalized L\"{u}scher transformations relevant for GWL symmetry 
on the lattice.
\medskip

\noindent {\bf (a)} First of all, we assume that $\D$ represents some 
acceptable lattice Dirac operator. By ``acceptable'' we mean the 
following: (a) correct classical continuum limit (b) locality (exponential 
decay at large distances) (c) invariance under symmetries of the hypercubic 
lattice (translations and symmetries of hypercube) (d) gauge invariance.
We will define the corresponding concepts precisely as we will need them
and denote the set of these acceptable operators as $\Dset$. Note 
that we do not include the absence of doublers here, which is convenient
to discuss separately.

Being composed of gauge fields, lattice Dirac operator actually 
represents a set of linear operators, one for every gauge configuration. 
We require the invariance of the fermionic action in arbitrary gauge 
background which results in the corresponding set of conditions 
(\ref{eq:20}). In this context, we will refer to them as GW relation. 
If $\Rb$ is trivial in spinor space, this reduces formally to the standard 
GW relation \cite{Gin82A}. 
\medskip

\noindent {\bf (b)} The aim is to interpret L\"{u}scher transformations
as generalized chiral transformations. However, in view of relation
(\ref{eq:30}), the corresponding symmetry of $\D$ neither poses
restriction on the set of acceptable operators, nor is it physically
interesting unless further requirements are imposed on the matrix $\Rb$.
Not surprisingly, the physically relevant restriction is given by the 
requirement that $\Rb$ be local \cite{Gin82A, Has98B}. The intuitive 
argument proceeds as follows: According to GW relation (\ref{eq:20}), 
$\Rb$ determines the character of the anticommutator of $\D^{-1}$ with 
$\gfive$. For $\Rb=0$, L\"{u}scher tranformations reduce to usual chiral 
transformations and chiral symmetry requires the propagator to anticommute 
with $\gfive$. Since the inherent feature of such lattice Dirac operators 
is doubling \cite{Nie81A}, we have to consider a nonzero $\Rb$. 
If $\Rb$ decays sufficiently fast, then propagator will anticommute with 
$\gfive$ at least 
at large distances which might still result in essentially chiral dynamics.
Indeed, as shown explicitly by Hasenfratz \cite{Has98B} in the context 
of standard GW relation, this is indeed true if $\Rb$ is local. 
We therefore restrict ourselves to {\it local nonzero} $\Rb$. 

In what follows, we will refer to $\D\in\Dset$ for which there exist 
a local nonzero $\Rb$ such that GW relation is satisfied as the operator 
with GWL symmetry (GW operator). To appreciate the power 
of this restriction, it is useful to consider the GW relation in the form 
(\ref{eq:30}) and to realize that $\D^{-1}$ is a non--local operator. 
For example, in the trivial gauge background ($U=1$),
the Fourier image of the propagator has the usual $1/p$ singularity. 
Such non--localities have to be canceled by $\F$ for 
arbitrary gauge configuration. 
Since $\F$ is chirally symmetric, this is possible if and
only if the non--locality of $\D^{-1}$ is entirely contained in its chirally
symmetric part. This is very restrictive on $\D$ and physically it
asserts that the chirally non--symmetric portion of the propagator does
not affect the long-distance physics at all. 
This is an essential property of GW operators that can be used 
as their alternative definition without any reference to operator $\Rb\,$: 
The set is defined by all $\D\in\Dset$ such that chirally nonsymmetric part 
of $\D^{-1}$ is local in arbitrary gauge background.

To make this explicit, we write $\D^{-1}$ in the relevant unique 
decomposition
\begin{equation}
    \D^{-1} \;=\; (\D^{-1})_C  + (\D^{-1})_N  \;,
    \label{eq:40}
\end{equation}
where $\{(\D^{-1})_C,\gfive\}=0$ and $[(\D^{-1})_N,\gfive]=0$. Then the
above discussion requires that $\F$ in relation (\ref{eq:30})
be written in the form $\F = -(\D^{-1})_C + \tilde{\F}$,
where $\tilde{\F}$ is arbitrary {\it local} chirally symmetric
matrix. Relation (3) then takes the form
\begin{equation}
   \Rb \;=\; (\D^{-1})_N + \tilde{\F}\qquad\qquad
   \{\tilde{\F},\gfive\}=0\qquad\quad 
   \tilde{\F}\; \mbox{local}\;.
   \label{eq:50}
\end{equation}

\noindent {\bf (c)} The final restriction is motivated by noting that 
according to the fundamental GW relation (\ref{eq:20}), adding a chirally
symmetric part to $\Rb$ has no effect on the dynamics dictated by the
GWL symmetry. We will therefore not reduce the set of GW
operators in any way if we only consider $\Rb$ whose chirally
symmetric part is identically equal to zero, i.e.
\begin{equation}
    [\,\Rb, \gfive\,] \;=\; 0\qquad\quad
    \mbox{or} \qquad\quad
    \Rb \;=\; \Rb_N
    \label{eq:60}
\end{equation}
Note that this restriction means seting $\tilde{\F}=0$ in relation
(\ref{eq:50}). 

In what follows, we will denote the set of all $\Rb$ that obey restrictions 
discussed in {\bf (b)} and {\bf (c)} as $\Rset$. It is the set of nonzero 
local $\Rb$, satisfying (\ref{eq:60}). For $\Rb\in\Rset$, the GW relation
(\ref{eq:20}) can be written in the form 
\begin{equation}
    \{\D,\gfive\} \;=\; 2\D\Rb\gfive\D \qquad\;
    \mbox{or} \qquad\;
    \Rb \;=\; (\D^{-1})_N\;.
    \label{eq:70}
\end{equation}
For future reference it is useful to assign to any $\D\in\Dset$, $\Rb\in\Rset$
an operator 
\begin{equation}
   \Dc \;\equiv\; 2\Rb\D\,,
   \label{eq:80}
\end{equation}
which brings the GW relation to the canonical form
\begin{equation}
    \{\Dc,\gfive\} \;=\; \Dc \gfive \Dc \qquad\;
    \mbox{or} \qquad\;
    \half\identity \;=\; (\Dc^{-1})_N\;.
    \label{eq:90}
\end{equation}
Here the first form is fundamental and the second one is equivalent
to it if $\Dc^{-1}$ can be meaningfully defined.
\smallskip

To summarize: In this subsection we have restricted the set of
pairs $(\D,\Rb)$ representing generalized L\"{u}scher transformations 
(\ref{eq:10}) to the subset where $\D\in\Dset$, $\Rb\in\Rset$ 
and the GW relation (\ref{eq:70}) is satisfied. We will denote the set 
of such transformations as $\Tset$. By construction, set $\Tset$ 
contains transformations physically relevant to the situation when GWL 
symmetry is present in the theory defined by acceptable lattice Dirac 
operator.

\subsection{The Statement of Main Result}

The main result of this paper can be expressed by the following
statement:

\bigskip
\noindent {\it Transformations contained in ${\frak T}$ 
couple infinitely many fermionic degrees of freedom on the
infinite lattice. Stated equivalently, these transformations
couple variables at arbitrarily large lattice distances, i.e. are
non-ultralocal.}
\bigskip

\noindent The above conclusion is based on the following considerations:

\begin{description}
  \item[$(\alpha)$] Because of the form of the generalized L\"{u}scher
  transformations, it is sufficient to show that the operator 
  $\Dc$, assigned to arbitrary $(\D,\Rb)\in {\frak T}$ in (\ref{eq:80}), 
  couples infinitely many fermionic degrees of freedom. 

  \item[$(\beta)$] We will prove the property of $\Dc$ required
  in $(\alpha)$ {\it rigorously} for free fermions, i.e. for the subset
  of generalized L\"{u}scher transformations, where gauge field is set
  to unity and the gauge--flavour structure is ignored. The flavour
  structure of $\D$ is trivial from the start and the gauge structure
  becomes so when $U=1$. GW relation (\ref{eq:70}) then enforces this 
  also on $\Rb$ and hence $\Dc$. 

  \item[$(\gamma)$] In gauge invariant theory, lattice
  sites coupled in trivial gauge background will also be coupled 
  in generic background. Hence, the same conclusion applies for this 
  case too.
\end{description}
\medskip

\noindent We stress that there are no physically interesting exceptions to 
the result formulated here. 

\section{Transformations in Unit Gauge Background}

In this section, we will consider the generalized L\"{u}scher 
transformations for free fermions. However, we will keep all the notation 
of the previous section and the restriction will be implicitly understood.
Since the gauge-flavour structure will be ignored, the operators 
considered here act on the vectors of $2^{d/2}$--component fermionic 
degrees of freedom living on the sites of an infinite hypercubic Euclidean 
lattice in $d$ even dimensions. Matrix $\G$ representing such operator can 
be uniquely expanded in the form
\begin{equation}
    \G_{m,n} = \sum_{a=1}^{2^d} \G^a_{m,n}\Gamma^a\;,
    \label{eq:100}
\end{equation}
where $m,n$ label the lattice points, $\G^a$ denotes a matrix with 
space--time indices, and $\Gamma^a$ is the element of the Clifford basis.
Clifford basis is built on gamma--matrices satisfying
$ \{ \gmu,\gnu \} = 2\dmunu\identity$. 
For example, in four dimensions we have  
$\Gamma\equiv\{\identity,\gmu,\gfive,\gfive\gmu,\sigma_{\mu\nu,(\mu<\nu)}\}$, 
where $\gfive=\gone\gtwo\gamma_3\gamma_4$,
$\sigma_{\mu\nu}\equiv {i\over 2}[\gmu,\gnu]$.
Because of the completeness of Clifford basis on the space of
$2^{d/2}\times 2^{d/2}$ complex matrices, Eq.~(\ref{eq:100}) describes 
arbitrary operator in question. In what follows we will refer to the
operators $\G^a$ as Clifford components of $\G$.

\subsection{Representation of Local Symmetric Operators} 

Since locality and invariance under symmetries of the hypercubic lattice
will play the crucial role in our discussion, we first define explicitly
the Fourier representation for operators that satisfy these requirements.
Hypercubic lattice structure is invariant under translations by arbitrary 
lattice vector and under the subgroup of $O(d)$ 
transformations -- hypercubic rotations and reflections. We refer to the 
former as {\it translation invariance} and to the latter as 
{\it hypercubic invariance}.

\begin{definition} 
(Locality) Operator $\G$ is said to be local if there are positive real 
constants $c$, $\delta$ such that all its Clifford components $\G^a$
satisfy
\begin{displaymath}
   |\G^a_{m,n}| \;<\; c\, e^{-\delta |m-n|}\qquad\quad
   \forall m,n.
\end{displaymath}
Here $|m-n|$ denotes the Euclidean norm of $m-n$.
\end{definition} 

\begin{definition} 
(Translation Invariance) Operator $\G$ is said to be translationally 
invariant if all its Clifford components $\G^a$ satisfy 
\begin{equation}
    \G_{m,n}^a = \G_{0,n-m}^a\equiv g_{n-m}^a \qquad\quad
    \forall m,n.
    \label{eq:110}
\end{equation}
\end{definition}

\begin{definition} 
(Hypercubic Invariance) Let $\Hcal$ be an element of the hypercubic group
in defining representation and $H$ the corresponding element of the
representation induced on hypercubic group by spinorial representation of 
$O(d)$. Operator $\G$ is said to have hypercubic invariance if for 
arbitrary $\Hcal$, $m$, $n$ we have
\begin{displaymath}
   \G_{n,m} \;=\; H^{-1} \G_{\Hcal n, \Hcal m}\, H\,.
\end{displaymath}
\end{definition}

Requirement of translation invariance and locality is equivalent to the 
existence of diagonal analytic Fourier images of space--time parts of $\G$. 
In particular 
\begin{equation}
    G^a(p)\equiv \sum_n g_n^a e^{ip\cdot n}\qquad\qquad
    G(p) \;\equiv\; \sum_{a=1}^{2^d} G^a(p)\Gamma^a\,,
    \label{eq:120}
\end{equation}
where functions $G^a(p)$ of lattice momenta $p\equiv (p_1,\ldots,p_d)$ 
are complex--valued, periodic and analytic. Adding hypercubic symmetry as 
an additional constraint, we now define explicitly the Fourier 
representation of local symmetric operators that we will use:
\begin{definition} (Set $\Gsl$)  
  Let $G^a(p),\; a=1,2,\ldots 2^d$, are the complex valued functions of real 
  variables $p_\mu$, and let $G(p)$ be the corresponding matrix function 
  constructed as in Eq.~{\rm (\ref{eq:120})}. We say that $G(p)$ belongs 
  to the set $\Gsl$ if:
     \begin{description}
       \item[$(\alpha)$] Every $G^a(p)$ is an anlytic function with 
       period $2\pi$ in all $p_\mu$.
       \item[$(\beta)$] For arbitrary hypercubic transformation $\Hcal$
       it is true identically that
       \begin{equation}
          G(p) \;=\; \sum_{a=1}^{2^d} G^a(p)\Gamma^a  
          \;=\; \sum_{a=1}^{2^d} G^a(\Hcal p) H^{-1}\Gamma^a H\,. 
          \label{eq:130}
       \end{equation}
     \end{description}
\end{definition}
We emphasize that the set $\Gsl$ is mathematically fully equivalent to the 
set of all local, translation invariant kernels $\G$. We can therefore speak
of $G(p)$ and $\G$ interchangeably and, indeed, we will frequently write 
$\G\in\Gsl$.
    
We finally note that since any hypercubic transformation $\Hcal$ can be 
decomposed into products of reflections of single axis (${\cal R}_\mu$) and 
exchanges of two different axis (${\cal X}_{\mu\nu}$), it is sufficient to 
require invariance under these operations. Transformation properties of all 
the elements of the Clifford basis are determined by the fact that $\gmu$
transforms as $p_\mu$ (vector). In particular
\begin{displaymath}
  R_\nu^{-1}\gmu R_\nu \;=\; \cases{-\gmu,& if $\mu=\nu$;\cr
                                     \gmu,& if $\mu\ne\nu$,\cr}
\end{displaymath}
and
\begin{displaymath}
  X_{\rho\sigma}^{-1}\gmu X_{\rho\sigma} \;=\;
                 \cases{\gamma_\sigma,& if $\mu=\rho$;\cr
                        \gamma_\rho,&   if $\mu=\sigma$;\cr
                        \gmu,&          otherwise,\cr}
\end{displaymath}
where $R_\mu, X_{\mu\nu}$ are the spinorial representations of 
${\cal R}_\mu, {\cal X}_{\mu\nu}$. The elements of the Clifford basis 
naturally split into groups with definite transformation properties and the 
hypercubic symmetry thus translates into definite algebraic requirements on 
functions $G^a(p)$ which we will later exploit.

\subsection{Sets $\Dset$, $\Rset$ and $\Tset$}

We now give the definition of fundamental sets that we introduced
in Sec.~2.

\begin{definition}
   (Set $\Dset$) Let $D(p)\in \Gsl$ be a local symmetric operator, 
   such that in the vicinity of $p=0$ its Clifford components $D^a(p)$ 
   satisfy
   \begin{equation}
      D^a(p) = 
        \cases{ip_\mu + O(p^2),&if $\Gamma^a  =\gmu\,$;\cr
        O(p^2),&if $\Gamma^a\ne\gmu\,,\forall\mu\,.$\cr}
      \label{eq:140}
   \end{equation}
   Collection $\Dset\subset\Gsl$ of such elements $D(p)$ defines 
   the set of acceptable lattice Dirac operators.
\end{definition} 

\begin{definition}
   (Set $\Rset$) Set $\Rset$ consists of all nonzero local operators $\Rb$
   such that condition {\rm (\ref{eq:60})} is satisfied.
\end{definition}

\begin{definition}
   (Set $\Tset$) We define $\Tset$ as the collection of pairs $(\D, \Rb)$ 
   such that $\D\in\Dset$, $\Rb\in\Rset$, and GW relation {\rm (\ref{eq:70})}
   is satisfied.
\end{definition}

The following simple auxiliary statement will be useful in what follows:
 
\begin{lemma}
   If $(\D,\Rb)\in\Tset$, then $\Rb\in\Gsl$ and $\Dc\equiv 2\Rb\D\in\Gsl$.
\end{lemma}

\begin{proof}
Since $\Rb\in\Rset$, it is local. GW relation (\ref{eq:70}) has to be
satisfied and from its second form it follows that $\Rb$ has to respect
symmetries of $\D$. Hence, $\Rb\in\Gsl$ and consequently $\Dc\in\Gsl$.
\end{proof}

\subsection{Ultralocality}

We now give a precise meaning to ultralocality and to the notion that 
operator couples ``infinitely many degrees of freedom''. By ultralocality 
we mean that the fermionic variables do not interact beyond some finite 
lattice distance: 
\begin{definition} (Ultralocality)
   Let ${\cal C}_N$ denotes the set of all lattice sites contained in the 
   hypercube of side $2N$, centered at $n=0$, i.e. 
   ${\cal C}_N\equiv\{n : |n_\mu|\le N,\, \mu=1,\ldots,d\}$. 
   Operator $\G$ is said to be be ultralocal if there is a positive integer 
   $N$, so that
   \begin{displaymath}
       \G^a_{m,n} \;=\; 0\,, \qquad\;          
       \forall m,n: (m-n) \not\in {\cal C}_N\,,\;\forall a\,.
   \end{displaymath} 
\end{definition}

When the ultralocal operator $\G$ acts on the vector of fermionic variables 
$\psi$, then every new $\psi^\prime_m = \G_{m,n}\psi_n$ is a linear
combination of finite number of variables residing in the corresponding 
hypercube ${\cal C}_N$ around point $m$. On the contrary, if the operator 
is non-ultralocal, then there exist a point $m$ such that the $\psi^\prime_m$
is a combination of infinite number of old variables. When $\G$ is 
translationally invariant, this is true for arbitrary point $m$. 

If operator $\G$ is translationally invariant and ultralocal, then
$\G\in\Gsl$ and for later reference it is useful to make explicit
the following simple statement:
\begin{lemma}
    Clifford components $G^a(p)$ coresponding to translation invariant
    ultralocal operator $G(p)$ are functions with finite number of
    Fourier terms.
\end{lemma}

\begin{proof}
This is a trivial consequence of ultralocality and the definition of 
Fourier image (\ref{eq:120}). In fact, ultralocality implies the
existence of ${\cal C}_N$ such that in the notation of Eq.~(\ref{eq:110}) 
we have
\begin{equation}
    G^a(p)\equiv \sum_{n\in{\cal C}_N} g_n^a e^{ip\cdot n}\;.
    \label{eq:150}
\end{equation}
\end{proof}

\subsection{Minimal Periodic Directions}

Consider straight lines in momentum space passing through the origin.
A special subset is defined by those lines for which all periodic 
functions $f(p)$ (periodic with $2\pi$ in all $p_\mu$) will remain 
periodic when restricted to that line. Such lines run, in addition
to origin, through other points $p$ such that 
$p_\mu=2\pi k_\mu, k_\mu\in\Z, \forall \mu$, and define so called 
periodic directions in momentum space. Periodic directions are special
from the point of view of hypercubic symmetry and symmetric functions
simplify on them accordingly. In this subsection, we will consider the 
subset of periodic directions (minimal periodic directions), for which 
the structure of elements in $\Gsl$ simplifies maximally when they are 
restricted to the corresponding lines.

\begin{definition} 
   (Restriction $\Delta^\rho$) Let $\rho \in \{1,2,\ldots d\}$ and let 
   $\pbar$ denotes the restriction of the momentum variable $p$ on the 
   line defined through 
   \begin{displaymath}
      \pbar_\mu \;=\; \cases{q,& if $\mu=1,\ldots,\rho$;\cr
                             0,& if $\mu=\rho+1,\ldots,d$.\cr}
   \end{displaymath}
   Map $\Delta^\rho$ that assigns to arbitrary function $f(p)$ of $d$ 
   real variables a function $\overline{f}(q)$ of single real
   variable through
   \begin{equation}
      \Delta^\rho\Bigl[\,f(p)\,\Bigr] \;\equiv\; 
      \overline{f}(q) \;\equiv\; f(\pbar)\,,
      \label{eq:160}
   \end{equation} 
   will be refered to as restriction $\Delta^\rho$. 
\end{definition}

The following auxiliary statement will be important in what follows:
\begin{lemma}
   Let $\rho \in \{1,2,\ldots d\}$. Let further $G(p)\in\Gsl$, and let
   $\Gr(q)$ be its restriction under $\Delta^\rho$ defined through 
   \begin{displaymath}
      \Gr(q) \;=\; \sum_{a=1}^{2^d} \Gr^a(q)\Gamma^a\qquad\quad  
      \Gr^a(q) \;=\; \Delta^\rho\Bigl[\,G^a(p)\,\Bigr]\,.
   \end{displaymath}
   Then $\Gr(q)$ can be written in form
   \begin{equation}
      \Gr(q) \;=\; X(q)\identity 
                 + Y(q)\sum_{\mu=1}^\rho \gmu\,,
      \label{eq:170}
   \end{equation}
   where $X(q)=X(-q)$, $Y(q)=-Y(-q)$ are analytic functions of one real 
   variable, periodic with $2\pi$.
\end{lemma}

\begin{proof}
Let's denote the following sets of indices for later convenience:
$u\equiv \{1,2\ldots d\}$, $u^\rho\equiv \{1,2\ldots \rho\}$.
It is useful to think of Clifford basis as subdivided into 
non--intersecting subsets
$\Gamma=\union_j \Gamma_{(j)}$, where $\Gamma_{(j)}$,
$j=0,1,\ldots,d$, contains the elements that can be written as the 
product of $j$ gamma-matrices. For example $\Gamma_{(0)}=\{\,\identity\,\}$,
$\Gamma_{(1)}=\{\,\gmu, \;\mu\in u\,\}$, and so on. With the 
appropriate convention on ordering of gamma--matrices in the definition 
of $\Gamma^a$, we can then rewrite the Clifford decomposition of $G(p)$ 
in the form
\begin{equation}
    G(p) \;=\; \sum_{j=0}^d \sum_{\mu_1,\mu_2\ldots\mu_j\atop
                                  \mu_1<\mu_2\ldots<\mu_j}
               F_{\mu_1,\mu_2\ldots\mu_j}(p)\, 
               \gamma_{\mu_1}\gamma_{\mu_2}\ldots\gamma_{\mu_j}\,,
    \label{eq:180}
\end{equation}
where all $\mu_i\in u$. We will now consider contributions to $\Gr(q)$ 
originating from different subsets $\Gamma_{(j)}$.

\smallskip
\noindent {\em (1) $j\ge2$} 
\smallskip

\noindent Consider arbitrary single term in decomposition (\ref{eq:180}), 
specified by the set of indices $v\equiv\{\mu_1,\mu_2\ldots\mu_j\}$. 
At least one of the following statements is true:

\begin{description}
   \item[(a)] Exists element $\mu\in v$, such that $\mu\notin u^\rho$.
   \item[(b)] Exist two elements $\mu,\nu\in v$ such that $\mu,\nu\in u^\rho$.
\end{description}  

\noindent Indeed, assume that both of the above statements are false. Then, 
from (a) it follows that $v\subset u^\rho$. Since (b) is also false, this 
means that $v$ contains at most one element. This is the contradiction with 
the assumption that $j\ge 2$.

If then (a) is true for our particular $v$, we can consider the
reflection ${\cal R}_\mu$ through the corresponding axis $\mu$. Since 
$
R_\mu^{-1} \,\gamma_{\mu_1}\gamma_{\mu_2}\ldots\gamma_{\mu_j}\, R_\mu
\,=\, -\gamma_{\mu_1}\gamma_{\mu_2}\ldots\gamma_{\mu_j}\,,
$
hypercubic symmetry of $G(p)$ requires
$
 F_{\mu_1,\mu_2\ldots\mu_j}({\cal R}_\mu p) =
 -F_{\mu_1,\mu_2\ldots\mu_j}(p)\,.
$
However, since $\mu\notin u^\rho$, the restricted variable $\pbar$ under 
$\Delta^\rho$ satisfies ${\cal R}_\mu \pbar = \pbar$, and hence 
\begin{displaymath}
   \Fr_{\mu_1,\mu_2\ldots\mu_j}(q) \;\equiv\;
   F_{\mu_1,\mu_2\ldots\mu_j}(\pbar) \;=\; -  
   F_{\mu_1,\mu_2\ldots\mu_j}(\pbar) \;=\; 0.
\end{displaymath}

Similarly, if (b) is true, we can apply the exchange ${\cal X}_{\mu\nu}$
of the axes $\mu,\nu\in u^\rho$. Then, again, the Clifford element 
is odd which forces this also on the corresponding function. However, 
$\pbar$ does not change under this operation and hence the restriction 
vanishes in this case too.

Consequently, $\Fr_{\mu_1,\mu_2\ldots\mu_j}(q)$ must vanish for any $v$.

\smallskip
\noindent {\em (2) $j<2$} 
\smallskip

\noindent After considerations of case {\em (1)}, we can write $\Gr(q)$
in the form
\begin{displaymath}
      \Gr(q) \;=\; X(q) \identity 
                 + \sum_{\mu=1}^d Y_\mu(q)\gmu\,,
\end{displaymath}
where $X(q),Y_\mu(q)$ are the restrictions of the corresponding 
Clifford elements. Invariance under reflection of the axis 
$\mu\notin u^\rho$ demands, however, that corresponding $Y_\mu(q)=0$. 
Moreover, if we exchange axes $\mu,\nu\in u^\rho$, then hypercubic 
symmetry implies
\begin{displaymath}
   Y(q) \;\equiv\; Y_1(q) \;=\; Y_2(q) \;=\;\ldots =\;Y_\rho(q)\,.
\end{displaymath}
This gives the desired form (\ref{eq:170}) and the reflection properties
of $X(q)$, $Y(q)$ follow from invariance under the product of reflections
${\cal R}_1{\cal R}_2\ldots{\cal R}_\rho$. Analyticity and periodicity are 
inherited from corresponding propereties of unrestricted operator.
\end{proof}

\subsection{Lemma}

The most important ingredient in the proof of our main theorem
will be the following auxiliary statement that was first formulated 
in Ref.~\cite{Hor98A}.

\begin{lemma}
  Let $K$,$L$ be nonnegative integers and $\rp$ a positive real number.
  Consider the set ${\cal F}^{K,L}$ of all pairs of functions 
  $(\,\A(q),\B(q)\,)$ that can be written in form
  \begin{equation}
     \A(q) = \sum_{-L\le n \le K} \fan  e^{iq\cdot n} \qquad\quad
     \B(q) = \sum_{-L\le n \le K} \fbn  e^{iq\cdot n} \;,
     \label{eq:app10}
  \end{equation}
  where $q\in\R ,\, n\in\Z$, and $\fan,\fbn\in\C$ are such that $\fa_K,\fb_K$
  do not vanish simultaneously and $\fa_{-L},\fb_{-L}$ do not vanish 
  simultaneously. Further, let ${\cal F}^{K,L}_\rp \subset {\cal F}^{K,L}$
  denotes the set of all solutions on ${\cal F}^{K,L}$ of the equation
  \begin{equation}
        \A(q)^2 + \rp\,\B(q)^2 = 1\;.
        \label{eq:app20}
  \end{equation} 
  Then the following holds:
    \begin{description}
      \item[$(\alpha)$] If $K=L=0$, then 
      $\;{\cal F}^{0,0}_\rp = 
      \{\,(\fa_0,\fb_0)\,:\, \fa_0^2 + \rp\,\fb_0^2=1\, \}$.

      \item[$(\beta)$] If $K=L>0$, then 
      $\;{\cal F}^{K,K}_\rp = \{\,(\,\A(q),\B(q)\,)\,\}$,
      such that 
        \begin{displaymath}
          \A(q) = \fa_{-K}\,e^{-iq\cdot K} + \fa_K\,e^{iq\cdot K} \qquad\qquad
          \B(q) = \fb_{-K}\,e^{-iq\cdot K} + \fb_K\,e^{iq\cdot K} 
        \end{displaymath}
      with  
        \begin{displaymath}
          \fb_K \ne 0 \qquad\quad 
          \fb_{-K} = {1\over {4\rp\,\fb_K}} \qquad\quad
          \fa_K = \,c\,i\sqrt{\rp}\,\fb_K \qquad\quad
          \fa_{-K} = {c\over {4i\sqrt{\rp}\,\fb_K}}
        \end{displaymath}
      where $\sqrt{\rp}>0$ and $c=\pm 1$.

      \item[$(\gamma)$] If $K\ne L$, then $\;{\cal F}^{K,L}_\rp = \emptyset$.
    \end{description} 
\end{lemma}

The usefulness of the above result lies in the fact that equation
(\ref{eq:app20}) arises as a GW condition (\ref{eq:90}) for $\Dc$ restricted
by $\Delta^\rho$. Lemma~4 provides us with classification of all solutions 
of this equation on the space of periodic functions with {\em finite} 
number of Fourier terms. Indeed, if both functions $A(q)$, $B(q)$ have only 
strictly positive (negative) Fourier components, then the equation clearly 
can not be satisfied. All other cases are covered by Lemma~4. Perhaps
surprisingly, the Fourier structure of solutions of 
Eq.~(\ref{eq:app20}) is thus either very simple (essentially a single 
Fourier component) or very complicated (infinitely many of them).
\bigskip

\begin{proof}
Case $(\alpha)$ of constant functions $A(q),B(q)$ is obvious and so we 
concentrate on cases $(\beta)$ and $(\gamma)$. Because of completeness and
othogonality of the Fourier basis, equation (\ref{eq:app20}) imposes the
following set of conditions on coefficients
$\fan,\fbn$ 
   \begin{equation}
     \sum_{\scriptstyle -L\le n\le K \atop
           \scriptstyle -L\le k-n\le K} \fa_n\fa_{k-n} \;+\;
     \rp\sum_{\scriptstyle -L\le n\le K \atop
            \scriptstyle -L\le k-n\le K} \fb_n\fb_{k-n} \;=\; \delta_{k,0} \;,
     \qquad -2L\le k \le 2K .
     \label{eq:app30}
   \end{equation} 

\vskip 0.16in
{\it Case $(\beta)$}
\vskip 0.16in

We split the set of equations (\ref{eq:app30}) into groups
that can be analyzed in sequence.

\smallskip
\noindent (I) $\;K\le k \le 2K$
\smallskip
 
\noindent Equations in this group involve the coefficients of nonnegative
frequencies only. Starting from $k=2K$ and continuing down we have
\footnote{For definitness of notation, we assume implicitly that $K$ is an 
even integer, but that distinction is only relevant for the notation, not
the argument.}
\begin{equation}  
   \begin{array}{llllll}
      &0\;=\;\fa_K^2 &+\quad \rp\,\fb_K^2 &&&\\
      &0\;=\;\fa_K\fa_{K-1} &+\quad \rp\,\fb_K\fb_{K-1} &&&\\
      &0\;=\;2\fa_K\fa_{K-2}+\fa_{K-1}^2 &+\quad
        \rp\,(2\fb_K\fb_{K-2}+ \fb_{K-1}^2) &&&\\ 
      &\vdots &&&&\\
      &0\;=\; 2\fa_K\fa_0 + 2\fa_{K-1}\fa_1 +\ldots + \fa_{K\over 2}^2 &+\quad
         \rp\,(2\fb_K\fb_0 + 2\fb_{K-1}\fb_1 +\ldots + \fb_{K\over 2}^2). &&& 
   \end{array}
   \label{eq:app40}    
\end{equation}
The first equation is equivalent to
\begin{equation}
    \fa_K = \,c\,i\sqrt{\rp}\,\fb_K \qquad\quad
    \sqrt{\rp}>0,\;\, c=\pm 1.
    \label{eq:app50} 
\end{equation}
Since $\fa_K,\fb_K$ are not simultaneously zero, it follows that they have
to be both nonzero. Inserting this into the second equation of
(\ref{eq:app40}) yields that also $\fa_{K-1} = \,c\,i\sqrt{\rp}\,\fb_{K-1}$.

This procedure can be repeated with the analogous result for other
coefficients. Indeed, a generic equation in this sequence has the 
schematic form
\begin{displaymath}
            2\fa_K\fa_{K-n} + f(\fa_{K-1},\fa_{K-2},\ldots,\fa_{K-n+1}) +    
   \rp\;\Bigl(2\fb_K\fb_{K-n} + f(\fb_{K-1},
             \fb_{K-2},\ldots,\fb_{K-n+1})\Bigr) = 0,
\end{displaymath}
where we have just grouped the variables conveniently. Since the relation
$\fa_j = \,c\,i\sqrt{\rp}\,\fb_j$ already holds for
$j=K,K-1\ldots,K-n+1$, the variables grouped by function $f$ will
drop out of the equation and we are left with 
$\fa_{K-n} = \,c\,i\sqrt{\rp}\,\fb_{K-n}$ as claimed. By induction, 
we thus have that the set of equations (\ref{eq:app40}) is equivalent to
\begin{equation}
    \fa_n = \,c\,i\sqrt{\rp}\,\fb_n \qquad\quad
    \sqrt{\rp}>0,\;\, c=\pm 1,\;\, n=0,1,\ldots,K.
    \label{eq:app60} 
\end{equation}   

\smallskip
\noindent (II) $\;-2K\le k\le -K$
\smallskip

\noindent We can use exactly the same reasoning for these equations as we
did for group (I) and transform them into
\begin{equation}
    \fa_{-n} = \,\cbar\,i\sqrt{\rp}\,\fb_{-n} \qquad\quad
    \sqrt{\rp}>0,\;\, \cbar =\pm 1,\;\, n=0,1,\ldots,K.
    \label{eq:app70} 
\end{equation}   
The constants $c,\cbar$ are related. To see that, we examine the equation
for $k=0$, namely
\begin{displaymath}
       \sum_{n=1}^K 2\fa_n\fa_{-n} + \fa_0^2  \;+\;
       \rp\,\Bigl(\,\sum_{n=1}^K 2\fb_n\fb_{-n} + \fb_0^2\,\Bigr) \;=\;1 
\end{displaymath}
Using equations (\ref{eq:app60}) and (\ref{eq:app70}) this reduces to
\begin{equation}
    (1-c\cbar)\,2\rp\sum_{n=1}^K \fb_n\fb_{-n} \;=\;1 \qquad
    \Longrightarrow \qquad \cbar=-c\,.
    \label{eq:app80}      
\end{equation}
Two useful implications of (\ref{eq:app60},\ref{eq:app70},\ref{eq:app80})
that we will use in examining the rest of the equations are
\begin{equation}
    \fa_0 \,=\,\fb_0 \,=\,0
    \label{eq:app90}       
\end{equation} 
and
\begin{equation}
   \fa_n\fa_m \,+\, \rp\,\fb_n\fb_m = \cases{0,&if $nm\ge0\,$;\cr
                            2\rp\,\fb_n\fb_m\,,&if $nm<0$.\cr}
   \label{eq:app100}
\end{equation}
 
Note that if $K=1$, we have no other equations available. (\ref{eq:app80})
reduces to $\fb_1\fb_{-1}=1/4\rp$, which together with (\ref{eq:app60}),
(\ref{eq:app70}) and (\ref{eq:app90}) imply the desired result. If $K>1$,
then we have groups of equations that mix the coefficients of positive
and negative frequencies.

\smallskip
\noindent (III) $\;1\le k\le K-1$
\smallskip

\noindent Because of constraints (\ref{eq:app100}), only the monomials that
are the products of one coefficient of positive frequency and one
coefficient of negative frequency will contribute. Starting from $k=K-1$
the equations are
\begin{eqnarray}  
   0 &=& \fb_K\fb_{-1} \nonumber \\
   0 &=& \fb_K\fb_{-2} \,+\, \fb_{K-1}\fb_{-1} \nonumber \\
   0 &=& \fb_K\fb_{-3} \,+\, \fb_{K-1}\fb_{-2} \,+\,
         \fb_{K-2}\fb_{-1} \nonumber \\
     &\vdots& \\
   0 &=& \fb_K\fb_{-(K-1)} \,+\, \fb_{K-1}\fb_{-(K-2)} \,+\ldots +\,
         \fb_2\fb_{-1}. \nonumber
   \label{eq:app110}    
\end{eqnarray}
Since $\fb_K\ne 0$, it follows from the first equation that $\fb_{-1}=0$.
Inserting this into the second equation we have $\fb_{-2}=0$, and by 
trivial induction 
\begin{equation}
   \fb_{-n} = 0 = \fa_{-n}    \qquad\qquad n=1,2,\ldots ,K-1\,,
   \label{eq:app120}
\end{equation}
where we have already used the result (\ref{eq:app70}).

\smallskip
\noindent (IV) $\;-K+1\le k\le -1$
\smallskip

\noindent Analogously to group (III), this set of equations combined
with result (\ref{eq:app60}) is equivalent to
\begin{equation}
   \fb_n = 0 = \fa_n   \qquad\quad n=1,2,\ldots ,K-1 \,.
   \label{eq:app130}
\end{equation}

\smallskip
\noindent (V) $\;k=0$
\smallskip

\noindent This is the only equation that is still available and we have 
already put it in the form (\ref{eq:app80}).
Using (\ref{eq:app120}) and (\ref{eq:app130}) this simplifies to
\begin{equation}
   \fb_K\fb_{-K} \,=\, {1\over{4\rp}}\,, 
   \label{eq:app140}
\end{equation}
which together with (\ref{eq:app60}),(\ref{eq:app70}) establishes the result
$(\beta)$.

\vskip 0.16in
{\it Case $(\gamma)$}
\vskip 0.16in

The strategy of splitting the total set of equations (\ref{eq:app30}) into
groups goes over to this case without any change (except for index ranges).

Assume first that $K>L$. If $L=0$, then all we have is a group (I)
of equations and the equation (V). In particular, result
(\ref{eq:app60}) implies $\fa_0^2 + \rp\,\fb_0^2 = 0$, while the equation
for $k=0$ reduces to $\fa_0^2 + \rp\,\fb_0^2 = 1$, thus leading to a
contradiction. If $L>0$, then we will also have groups (II) and
(III). However, since $L<K$, the result (\ref{eq:app120}) implies that
coefficients of all negative frequencies now vanish $\fa_{-n}=\fb_{-n}=0$,
for ($n=1,2,\ldots L$). Consequently, equation for $k=0$ again reduces to 
$\fa_0^2 + \rp\,\fb_0^2 = 1$, which contradicts (\ref{eq:app60}) and there is
no solution.

For $K<L$ the same line of logic leads to the same conclusion, which
completes the proof.
\end{proof} 

\subsection{Theorem}

Required tools are now in place to prove the following theorem:

\begin{theorem}
If $(\D,\Rb)\in \Tset$, then $\Dc = 2\Rb\D$ is not ultralocal. 
\end{theorem}

\begin{proof}
We will proceed by contradiction. Let us therefore assume that
there exist $(\D,\Rb)\in \Tset$ such that $\Dc$ actually is ultralocal
and the following steps will lead us to contradiction:

$(\alpha)$ According to Lemma~1, $\Rb\in\Gsl$. Consequently, its
Clifford components are analytic with well defined Taylor series. 
In particular, let us for later convenience write explicitly
\begin{equation}
   R^a(p) \;=\; {r\over 2} + O(p)\qquad
   \mbox{if}
   \qquad \Gamma^a = \identity\,.
   \label{eq:190}     
\end{equation}  
 
$(\beta)$ We now consider the restriction $\Dcr(q)=2\Rbr(q)\Dr(q)$ under
$\Delta^\rho$. Taking into account Lemma~3, local properties 
(\ref{eq:140}) of $D(p)$, the fact that $[R,\gfive]=0$, and using notation 
of (\ref{eq:190}), we can conclude:
\begin{equation}
      \Dcr(q) \;=\; \Bigl( 1- A(q) \Bigr) \identity 
                 + i B(q)\sum_{\mu=1}^\rho \gmu\,,
      \label{eq:200}
\end{equation}
where $A(q)$, $B(q)$ are analytic functions periodic with $2\pi$, 
such that following properties around $q=0$
are satisfied\footnote{Note that we are not 
strict about enforcing all the consequences of hypercubic symmetry because 
it is not necessary. For example, one can easily see that hypercubic 
symmetry requires 
the Taylor reminder in Eq.~(\ref{eq:190}) be actually $O(p^2)$, and
the reminder of $B(q)$ in Eq.~(\ref{eq:210}) be $O(q^3)$.}
\begin{equation}
   A(q) = 1 + O(q^2) \qquad\quad B(q) = rq + O(q^2).
   \label{eq:210}
\end{equation}

$(\gamma)$ GW relation for $\Dcr$ given in Eq.~(\ref{eq:200}) takes a simple
form
\begin{equation}
    A(q)^2 + \rho\,B(q)^2 = 1,
   \label{eq:220}
\end{equation}
and, according to Lemma~2, ultralocality of $\Dc$ implies that
$A(q),B(q)$ have Fourier series with finite number of terms. 

$(\delta)$ Because of $(\gamma)$, we can apply Lemma~4 to conclude
that functions $A(q),B(q)$ must either be the constants, or there
is an integer $K_\rho>0$, such that 
\begin{displaymath}
    A(q) = a_{-K_{\rho}} e^{-iq\cdot K_\rho} 
         + a_{K_\rho} e^{iq\cdot K_\rho}\qquad\quad
    B(q) = b_{-K_{\rho}} e^{-iq\cdot K_\rho} 
         + b_{K_\rho} e^{iq\cdot K_\rho}\,.
\end{displaymath}
Local properties (\ref{eq:210}) exclude the constants, while in the
second case they dictate that the solutions are $A(q)=\cos(K_\rho q)$,
$B(q)=r\sin(K_\rho q)/K_\rho$. For these functions we have
\begin{equation}
   A^2 + \rho\,B^2 \;\,=\;\, 
   \cos^2(K_\rho q) \;+\; {{r^2\,\rho} \over {K_\rho^2}}\,\sin^2(K_\rho q)\,.
   \label{eq:230}
\end{equation}

$(\epsilon)$ In view of Eqs.~(\ref{eq:220},\ref{eq:230}) we have to 
distinguish two cases:
\medskip

(a) If $r=0$, Eq.~(\ref{eq:220}) can not be identically satisfied and we
already have a contradiction.
\medskip

(b) If $r\ne 0$, then to satisfy Eq.~(\ref{eq:220}) we have to
demand
\begin{displaymath}
      r \;=\; c\, {K_\rho \over \sqrt{\rho}} \qquad\quad 
      c=\pm 1,\;\, \sqrt{\rho}>0\,.
\end{displaymath}
This condition has to be satisfied for all $\rho\in\{1,2,\ldots d\}$.
In particular, if $\rho$ is a square of another integer 
($\rho=1$, for example), we have to conclude that $r$ is a rational number.
At the same time, if $\rho$ is not a square of an integer 
($\rho=2$, for example) we have to conclude that $r$ is irational and we
thus have a contradiction for $r\ne 0$ as well. This completes our proof. 

\end{proof}

The above result implies that every transformation (\ref{eq:10})
corresponding to $\D\in\Dset$ with GWL symmetry couples variables 
at arbitrarily large lattice distances. Every transformed
variable is a linear combination of infinitely many original ones.
This establishes the {\em weak theorem on ultralocality} for
GWL symmetry.

\subsection{Ultralocality of Symmetric Actions} 

Theorem~1 has the following useful immediate consequence:

\begin{corollary}
   If $(\D,\Rb)\in \Tset$ and $\Rb$ is ultralocal, then $\D$ must be
   non-ultralocal.
\end{corollary}

\noindent In other words, the lattice Dirac operator $\D\in\Dset$, satisfying 
GW relation (\ref{eq:70}) with ultralocal $\Rb$ can not be 
ultralocal~\cite{Hor98A}.\footnote{
For $\Rb$ trivial in spinor space, this result was stated in 
Ref.~\cite{Hor98A} as simple extension of canonical case by techniques 
discussed there. The proof was presented for example at VIELAT98 workshop. 
Shortly before this paper was ready for release, W.~Bietenholz posted a 
note~\cite{Bie99A}, where he uses these techniques in a similar 
fashion. Contrary to the original statement in that note, its revised version
appears to claim the case identical to one discussed in Ref.~\cite{Hor98A} 
and here.}
This has some unfortunate drawbacks for practical use of actions in this
category: It complicates perturbation theory, one looses obvious
numerical advantages steming from sparcity of the conventional operators,
and the question of simulating them is nontrivial and widely open. 
Moreover, while locality can be ensured easily for free case, it is usually 
not obvious in the presence of gauge fields if the action is not 
ultralocal. Studies such as~\cite{Her98A} will probably be necessary for 
any individual operator that might be of interest.

We stress that no definite conclusion on ultralocality of $\D$ from
Theorem~1 can be made if corresponding $\Rb$ is not ultralocal. In fact, 
there exist elements $(\D,\Rb)\in\Tset$ such that $\D$ is ultralocal when
$\Rb$ is not. For example, we can take
\begin{equation}
   D(p) \;=\; \Bigl(\sum_{\mu=1}^d \sin^2p_\mu\Bigr)\, \identity 
        \,+\, i\sum_{\mu=1}^d \sin p_\mu\,\gmu\,,
   \label{eq:240}
\end{equation}
which satisfies GW relation with 
$R(p)=\identity/(1+\sum_{\mu=1}^d \sin^2p_\mu)$. The point is that set 
$\Dset$ also contains operators with doublers and the above example is
one of them. Theorem~1 and Corollary~1 are valid regardless of whether
the action is doubler--free or not. However, the hypothesis on the
absence of ultralocal GW actions at free level can only hold if operators 
with doublers are excluded as they should. In view of our discussion 
leading to GW relation (\ref{eq:70}), one could prove the {\em strong 
theorem on ultralocality} by proving the following hypothesis:

\begin{hypothesis}
   There is no $D(p)\in \Dset$ such that the following three
   requirements are satisfied simultaneously:	
   \begin{description}
     \item[$(\alpha)$] $D(p)$ involves finite number of Fourier terms. 
     \item[$(\beta)$] $\Bigl(D^{-1}(p)\Bigr)_N$ is analytic.
     \item[$(\gamma)$] $\Bigl(D^{-1}(p)\Bigr)_C$ is analytic except if
                       $\,p_\mu=0\!\pmod{2\pi},\, \forall\mu$. 
   \end{description} 
\end{hypothesis}

\noindent Needless to say, it would be rather interesting to have a 
rigorous answer to whether the above hypothesis holds or not.
 
\section{Conclusion}

The long--standing quest for incorporating chiral fermionic dynamics on the 
lattice properly, culminated recently in the construction of natural 
field--theoretical framework for studying questions related to this 
issue. Central building block of this framework is the notion of 
L\"{u}scher transformations and corresponding GWL symmetry. 
While standard chiral transformation appears to be a smooth limiting
case of generalized L\"{u}scher transformations (\ref{eq:10}), we 
argue here that there is a sharp discontinuity in the behaviour of the two 
cases when the underlying fermionic dynamics exhibits the corresponding 
symmetry. While chiral transformation only mixes variables on a single
site, infinitesimal GWL symmetry operation allways requires rearrangement 
of infinitely many degrees of freedom and couples variables at arbitrarily 
large lattice distances. 

The above discontinuity is apparently at the heart of the fact that while
fermion doubling is a definite property of chiral symmetry, it is an 
indefinite property of GWL symmetry. At the same time, luckily, dynamical 
consequences are not affected by this discontinuity. This appears  
to support the general picture which says that imposing a proper chiral 
dynamics without doubling requires a delicate cooperation of many fermionic
degrees of freedom. These have to conspire to ensure that the chirally
nonsymetric part of the action does not affect the long distance 
behaviour of the propagator, and that the would--be doublers from the chirally
symmetric part become heavy.

Our discussion assumes that acceptable fermionic actions respect
symmetries of the hypercubic lattice structure. This is reasonable since
it guarantees the recovery of the corresponding Poincar\'e symmetries
of Minkowski space in the continuum limit without tuning. We rely
quite heavily on the consequences of hypercubic symmetry in 
particular, and so it would be interesting to know whether the picture 
changes if only the translation invariance is retained. At free
level, there indeed is a difference here for there exist ultralocal
L\"{u}scher transformations with symmetric lattice Dirac kernels.
For example, in two dimensions we can consider the ultralocal operator
\begin{equation}
   D(p) \;=\; (1-\cos p_1\cos p_2)\,\identity 
        \,+\, i\sin p_1\cos p_2\,\gone \,+\, i\sin p_2\,\gtwo\,,
   \label{eq:250}
\end{equation}
which does not respect hypercubic symmetry, satisfies GW relation
(\ref{eq:70}) with $R(p)=1/2$, and the L\"uscher transformation is 
ultralocal. However, there is a doubler at $p=(\pi,\pi)$. 
Therefore, we can only hypothesize that if the requirement of
hypercubic symmetry is lifted, there are no ultralocal L\"uscher 
transformations involving doubler--free $D(p)$. It would
be interesting to clarify whether that is indeed the case and also
whether non--ultralocality of symmetric doubler--free actions holds.

\bigskip
{\bf Acknowledgement:} I thank H.~Thacker for many pleasant discussions
on the issues discussed here, and to V.~Balek for useful input on the
case without hypercubic symmetry.

\end{document}
\bye